\documentclass[preprint,aps,prd,eqsecnum,superscriptaddress]{revtex4}
\usepackage[T1]{fontenc}
\usepackage[latin1]{inputenc}
\usepackage{graphicx,amsfonts}
\usepackage{amsmath}
\newcommand{\be}{\begin{equation}}
\newcommand{\ee}{\end{equation}}
\newcommand{\bea}{\begin{eqnarray}}
\newcommand{\eea}{\end{eqnarray}}
\newcommand{\bml}{\begin{subequations}}
\newcommand{\eml}{\end{subequations}}

\newcommand{\vx}{\vec{x}}

\newcommand{\vk}{\vec{k}}

\newcommand{\vp}{\vec{p}}
\newcommand{\vq}{\vec{q}}

\newcommand{\ep}{\epsilon}

\newcommand{\op}{\mathcal{O}}
\newcommand{\opt}{\tilde{\mathcal{O}}}

\begin{document}
\title{Fourth order spatial derivative gravity}
\author{F. S. Bemfica}
\email{fbemfica@fma.if.usp.br}
\affiliation{Instituto de Física, Universidade de São Paulo\\
Caixa Postal 66318, 05315-970, São Paulo, SP, Brazil}
\author{M. Gomes}
\email{mgomes@fma.if.usp.br}
\affiliation{Instituto de Física, Universidade de São Paulo\\
Caixa Postal 66318, 05315-970, São Paulo, SP, Brazil}
\date{\today}

\begin{abstract}
In this work we study a modified theory of gravity that contains up to fourth order spatial derivatives as a model for the Ho\v rava-Lifshitz gravity. The propagator is evaluated and, as a result, it is obtained one extra pole corresponding to a spin two nonrelativistic massless particle, an extra term which jeopardizes renormalizability, besides the unexpected general relativity  unmodified propagator. Then, unitarity is proved at the tree-level, where the general relativity pole has shown to have no dynamics, remaining only the two degrees of freedom of the new pole. Next, the nonrelativistic effective potential is determined from a scattering process of two identical massive gravitationally interacting bosons. In this limit, Newton's potential is obtained, together with a Darwin-like term that comes from the extra non-pole term in the propagator. Regarding renormalizability, this extra term may be harmful, by power counting, but it can be  eliminated by adjusting the free parameters of the model. This adjustment is in accord with the detailed balance condition suggested in the literature and  shows that the way in which extra spatial derivative terms are added is of fundamental importance.
\end{abstract}

\maketitle

\section{Introduction}

Einstein's General Relativity  (GR) generalizations due to the addition of extra derivative terms was proposed a long time ago~\cite{Weyl1,Eddington}. Such modifications became interesting in the context of quantum gravity, where Einstein's theory  is known to be perturbatively nonrenormalizable  by power counting \cite{Weinberg2}.  It has been verified that, in the presence of extra terms containing products of the curvature tensor, the theory turned out to be renormalizable. However, such modification  introduced pathologies into the theory which loses unitarity~\cite{Stelle}.

More recently, in an attempt of constructing a perturbatively renormalizable and unitary gravitational theory, Ho\v rava~\cite{Horava2} proposed modifications of GR via extra terms with only spatial derivatives, introduced in a chosen spacetime foliation. The foremost argument for such proposal lies in the fact that, the gravity propagator of the linearized theory would behave like
\be
\label{1}
\frac{1}{\omega^2-\vk^2-a_2(\vk^2)^2-\cdots-a_z(\vk^2)^z}\,,
\ee
$a_{2},\,\cdots,\,a_z$ being coupling constants, $k^\mu=(\omega,\vk)$ the four-momentum of the graviton and $z>1$ a parameter associated with the highest order of spatial derivatives. The absence of higher order time derivatives may transform the propagator as the one shown in (\ref{1}), with only simple poles in $\omega^2$. It has been verified that the theory proposed in~\cite{Horava2} is not so simple and that new degrees of freedom, among other illness, are present~\cite{Sotiriou,Henneaux,Pons,Bellorin,Padilla1,Padilla2,Padilla3,Bogdanos,Chaichian3}.

In this work we will be interested to study the exact form for (\ref{1}). To this end, we propose a prototype for the theory in~\cite{Horava2}. The model in question, restricting ourselves to $z=2$, is described by the action
\be
\label{2}
S=\frac{1}{\kappa^2}\int_\Re dt\int_\sigma d^3x \sqrt{-g}\left(R^{(4)}+\alpha R^{(3)2}+\beta R^{(3)ij}R^{(3)}_{ij}\right)\,,
\ee
defined in the foliation $\mathcal{M}\cong \Re\times\sigma$. There, spacetime indices $\mu,\,\nu,\,\cdots$ run from $0$ to $3$, while $i,\,j=1,2,3$ are the indices that label spacial coordinate on $\sigma$. Also, $\kappa^2=16\pi G$, $G$ is the Newton's constant, and, all over the paper, $c=\hbar=1$. The spacetime signature we are dealing with is $(-+++)$, $R^{(4)\alpha}_{\;\;\mu\nu\beta}$ ($R^{(3)^i}_{\;\;jkl}$) is the Riemann curvature tensor in 4 (3) dimensions for the metric $g_{\mu\nu}$ ($g_{ij}$). It will be convenient, in this spacetime signature, to define $R^{(4)\alpha}_{\;\;\mu\beta\nu}=\partial_\beta \Gamma^{(4)\alpha}_{\mu\nu}-\partial_\mu\Gamma^{(4)\alpha}_{\beta\nu}+\cdots$, while  $R^{(4)}_{\mu\nu}=R^{(4)\beta}_{\;\;\mu\beta\nu}$. Similar definitions are valid for the 3-dimensional curvature tensor.

Notice that, to be as general as possible, we could have incorporated the term $R^{(3)ijkl}R^{(3)}_{ijkl}$ into (\ref{2}). However, in the 3-dimensional case the Weyl tensor \cite{Weyl1} is identically zero. As a consequence, the aforementioned term $R^{(3)ijkl}R^{(3)}_{ijkl}$ can be totaly cast as a combination of $R^{(3)2}$ and $R^{(3)ij}R^{(3)}_{ij}$ (see, for instance, the appendix in \cite{Schmidt}).

In the next section we will dedicate ourselves to the computation of the propagator of (\ref{2}) in the weak field approximation. Section \ref{sec:3} contains a systematic study of unitarity at the tree-level. At this level, we show that the dynamic of the theory is governed by a pole that corresponds to a nonrelativistic spin two massless particle. The obtention of the semiclassical nonrelativistic potential for a boson-boson scattering process, via gravitational interaction, is done in Section \ref{sec:4}. Section \ref{Conclusion} contains the conclusions.

\section{The propagator}
\label{sec:2}

Let us perform the linearization of (\ref{2}) in the weak field approximation
\be
\label{2.1}
g_{\mu\nu}\approx \eta_{\mu\nu}+\kappa h_{\mu\nu},
\ee
whose inverse is $g^{\mu\nu}\approx\eta^{\mu\nu}-\kappa h^{\mu\nu}$. The background metric is $\eta_{\mu\nu}=diag(-1,1,1,1)$ and the  $h_{\mu\nu}$ are the gravitational field fluctuations. By collecting  terms up to second order in $h_{\mu\nu}$, (\ref{2}) gives
\be
\label{2.2}
\underline{\underline{S}}=\frac{1}{2}\int d^4x \left[\frac{1}{2}h_{\mu\nu}\partial^2 h^{\mu\nu}-\frac{1}{4}h\partial^2 h+ \underline{\Gamma}_\mu\underline{\Gamma}^\mu+2\alpha\underline{R}^{(3)2}+2\beta\underline{R}^{(3)ij}\underline{R}^{(3)}_{ij}\right]\,,
\ee
where $\partial^2=\partial^\mu\partial_\mu$, simple underline means first order in $h$, double underline means second order in $h$ and so on. The trace is $h\equiv\eta^{\mu\nu}h_{\mu\nu}$ while
\bea
\label{2.3}
\underline{\Gamma}^\mu&\equiv&\eta^{\alpha\beta}{\underline{\Gamma}}^\mu_{\alpha\beta}\nonumber\\
&=&\partial_\alpha h^{\mu\alpha}-\frac{1}{2}\partial^\mu h\,.
\eea

Clearly, the action (\ref{2.2}) is invariant under the gauge transformations $\delta h_{\mu\nu}=-\partial_\mu\xi_\nu-\partial_\nu\xi_\mu\equiv-2\partial_{(\mu}\xi_{\nu)}$ for any arbitrary $\xi_\mu$. Then, in order to evaluate the propagator, we may choose the de Donder  gauge, $\underline{\Gamma}_\mu=0$, by introducing into the action the following term:
\be
\label{2.4}
S_{gf}=-\frac{\lambda}{2}\int d^4x \underline{\Gamma}_\mu\underline{\Gamma}^\mu\,.
\ee
Now, we may rewrite (\ref{2.2}) as
\be
\label{2.5}
\underline{\underline{S}}_{\lambda}= \underline{\underline{S}}+ S_{gf}=\frac{1}{2}\int d^4x h^{\mu\nu}\op_{\mu\nu,\alpha\beta}h^{\alpha\beta}\,,
\ee
where the operator $\op$ possesses the symmetries $\op_{\mu\nu,\alpha\beta}=\op_{\alpha\beta,\mu\nu}=\op_{\nu\mu,\alpha\beta}$. It is convenient to separate $\op$ into
\be
\label{2.6}
\op_{\mu\nu,\alpha\beta}=\op_{\mu\nu,\alpha\beta}^1+\op_{\mu\nu,\alpha\beta}^2\,,
\ee
where in momentum space, with  $k^\mu=(\omega,\vk)$,
\bea
\opt_{\mu\nu,\alpha\beta}^1&\equiv&-\frac{k^2}{2}\delta_{\mu\nu,\alpha\beta}+\frac{2-\lambda}{4}k^2\eta_{\mu\nu}\eta_{\alpha\beta}
+(1-\lambda)\eta_{((\mu(\nu}k_{\beta)}k_{\nu))}\nonumber\\
&&-\frac{1-\lambda}{2}\left(\eta_{\mu\nu}k_\alpha k_\beta +\eta_{\alpha\beta}k_\mu k_\nu\right)\,,\label{2.7a}
\eea
whereas $\opt_{\mu\nu,\alpha\beta}^2$ is equal to zero whenever one of its indices is time-like and
\bea
\opt_{ij,kl}^2&=&
\frac{\beta}{2}\vk^4\delta_{ij,kl}+(2\alpha+\beta)k_ik_jk_lk_k+(-2\alpha-\frac{\beta}{2})\vk^2\left(\delta_{ij}k_kk_l+\delta_{kl}k_ik_j\right)\nonumber\\
&&+(2\alpha+\frac{\beta}{2})\vk^4\delta_{ij}\delta_{kl}
-\beta\vk^2\delta_{((i(k}k_{l)}k_{j))}\,.\label{2.7b}
\eea
In the above expressions we have set $k^2=-\omega^2+\vk^2$, $\delta_{\mu\nu,\alpha\beta}\equiv \eta_{\mu(\alpha}\eta_{\beta)\nu}$ and used the convention $A_{((\mu(\alpha}B_{\beta)\nu))}=(1/2)(A_{\mu(\alpha}B_{\beta)\nu}+A_{\nu(\alpha}B_{\beta)\mu})$.

The difficulty of inverting (\ref{2.6}) comes from the noncovariant form of the operator $\opt$. Our plan is to first separate the components with pure spatial indices  from the rest. This may be achieved by defining
\bml
\label{2.8}
\bea
A_{\mu\nu,\alpha\beta}&\equiv&\opt_{\mu\nu,\alpha\beta}^1-\delta_{\mu\nu}^{ij}\opt_{ij,kl}^1\delta_{\alpha\beta}^{kl}\label{2.8a}\\
B_{ij,kl}&\equiv&\opt_{ij,kl}^1+\opt_{ij,kl}^2\,.
\eea
\eml
Componentwise,
\bml
\label{2.9}
\bea
A_{00,00}&=&-\frac{\lambda}{4}k^2\,,\label{2.9a}\\
A_{00,ij}&=&\frac{1}{2}\delta_{ij}\left[\frac{\lambda}{2}k^2-k^2-(1-\lambda)\omega^2\right]+\frac{1-\lambda}{2}k_ik_j\,,\label{2.9b}\\
A_{0i,kl}&=&\frac{1-\lambda}{2}\omega\left(-\delta_{i(k}k_{l)}+ k_i\delta_{kl}\right)\,,\label{2.9c}\\
M_{ij}&\equiv&A_{0i,0j}\nonumber\\
&=&\frac{1}{4}\delta_{ij}\left[k^2+(1-\lambda)\omega^2\right]-\frac{1-\lambda}{4}k_ik_j\,,\label{2.9d}
\eea
\eml
and zero otherwise. We now may write
\be
\label{2.9-2}
\opt_{\mu\nu,\alpha\beta}=A_{\mu\nu,\alpha\beta}+\delta_{\mu\nu}^{ij}B_{ij,kl}\delta^{kl}_{\alpha\beta}.
\ee

From the inverse equation
\be
\label{2.10}
\opt_{\mu\nu,\alpha\beta}\opt^{-1\,\alpha\beta,\lambda\sigma}=\delta_{\mu\nu}^{\lambda\sigma}\,,
\ee
one obtain
\bml
\label{2.11}
\bea
\opt^{-1\,00,00}&=&\frac{1}{A_{00,00}}\left(1-A_{00,ij}\opt^{-1\,ij,00}\right)\,,\label{2.11a}\\
\opt^{-1\,00,mn}&=&-\frac{A_{00,kl}}{A_{00,00}}\opt^{-1\,kl,mn}\,,\label{2.11b}\\
\opt^{-1\,0i,00}&=&-\frac{1}{2}M^{-1\,iq}A_{0q,kl}\opt^{-1\,kl,00}\,,\label{2.11c}\\
\opt^{-1\,0i,mn}&=&-\frac{1}{2}M^{-1\,ij}A_{0j,kl}\opt^{-1\,kl,mn}\,,\label{2.11cc}\\
C_{ij,kl}\opt^{-1\,kl,mn}&=&\delta^{mn}_{ij}\,.\label{2.11d}
\eea
\eml
By the definition of $M_{ij}$ in (\ref{2.9d}), it is straightforward to get
\be
\label{2.12}
M^{-1\,ij}=\frac{4}{k^2+(1-\lambda)\omega^2}\left(\delta^{ij}+\frac{1-\lambda}{\lambda k^2}k^ik^j\right)\,.
\ee
In the last line of (\ref{2.11}) we defined
\bea
\label{2.13}
&&C_{ij,kl}\equiv-\frac{A_{00,kl}A_{ij,00}}{A_{00,00}}-A_{ij,0m}M^{-1\,mn}A_{0n,kl}+B_{ij,kl}\nonumber\\
&&=\delta_{ij,kl}\left(-\frac{k^2}{2}+\frac{\beta}{2}\vk^4\right)\nonumber\\
&&+\frac{k_ik_jk_lk_k}{\vk^4}\left[(2\alpha+\beta)\vk^4+\frac{(1-\lambda)^2\vk^4}{\lambda k^2}-\frac{(1-\lambda)^3\omega^2\vk^4}{\lambda k^2[k^2+(1-\lambda)\omega^2]}\right]\nonumber\\
&&+\frac{\delta_{ij}k_kk_l+\delta_{kl}k_ik_j}{\vk^2}\left[-(2\alpha+\frac{\beta}{2})\vk^4-\frac{1-\lambda}{2}\vk^2
-\frac{1-\lambda}{\lambda k^2}\left(\frac{2-\lambda}{2}k^2+(1-\lambda)\omega^2\right)\vk^2\right.\nonumber\\
&&\left.+\frac{(1-\lambda)^2\omega^2\vk^2}{\lambda k^2}\right]
+\delta_{ij}\delta_{kl}\left[(2\alpha+\frac{\beta}{2})\vk^4+\frac{2-\lambda}{4}k^2+\frac{1}{\lambda k^2}\left(\frac{2-\lambda}{2}k^2+(1-\lambda)\omega^2\right)^2\right.\nonumber\\
&&\left.-\frac{(1-\lambda)^2\omega^2\vk^2}{\lambda k^2} \right]
+\frac{4\delta_{((i(k}k_{l)}k_{j))}}{\vk^2}\left[-\frac{\beta}{4}\vk^4+\frac{1-\lambda}{4}\vk^2
-\frac{(1-\lambda)^2\omega^2\vk^2}{4[k^2+(1-\lambda)\omega^2]}\right]\,.
\eea

The appendix \ref{ApendiceA} contains all the necessary tools to invert $C_{ij,kl}$. The calculation is tedious but straightforward. We use the Barnes-Rivers operators given in (\ref{23}), together with (\ref{25-1}), (\ref{25-2}), and, as a last step in the equation so obtained, the equalities in (\ref{28}) and (\ref{29}) are employed. We then quote the result
\bea
\label{2.14}
&&\opt^{-1}_{ij,kl}=\frac{\vk^2}{-\lambda k^4}\left(2P^1+4\bar{P}^0\right)_{ij,kl}\nonumber\\
&&+\frac{1}{-k^2}\left(-\frac{2\omega^2}{k^2}P^1+\frac{k^2-4\vk^2}{k^2}\bar{P}^0-\bar{\bar{P}}^0\right)_{ij,kl}\nonumber\\
&&+\frac{2P^2_{ij,kl}}{-k^2+\beta \vk^4}-(8\alpha+3\beta)\frac{\vk^4}{k^4}\bar{P}^0\nonumber\\
&&=\frac{\left(2\mathcal{P}^1+4\bar{\mathcal{P}}^0\right)_{ij,kl}}{-\lambda k^2}+\frac{\left(2\mathcal{P}^2-\mathcal{P}^0
-3\bar{\mathcal{P}}^0-\bar{\bar{\mathcal{P}}}^0\right)_{ij,kl}}{-k^2}\nonumber\\
&&+2P^2_{ij,kl}\left(\frac{1}{-k^2+\beta\vk^4}-\frac{1}{-k^2}\right)-(8\alpha+3\beta)\bar{\mathcal{P}}^0_{ij,kl}\,.\nonumber\\
\eea
So far we have obtained the part of the propagator with pure spatial indices. Equations (\ref{2.11a}), (\ref{2.11b}), (\ref{2.11c}) and (\ref{2.11cc}) enable us to get the remaining terms, so that
\bea
\label{2.15}
\opt^{-1}_{\mu\nu,\alpha\beta}
&=&\left(\frac{2\mathcal{P}^1+4\bar{\mathcal{P}}^0}{-\lambda k^2}+\frac{2\mathcal{P}^2-\mathcal{P}^0
-3\bar{\mathcal{P}}^0-\bar{\bar{\mathcal{P}}}^0}{-k^2}\right)_{\mu\nu,\alpha\beta}\nonumber\\
&&+\delta^{ij}_{\mu\nu}\delta^{kl}_{\alpha\beta}2P^2_{ij,kl}\left(\frac{1}{-k^2+\beta\vk^4}-\frac{1}{-k^2}\right)\nonumber\\
&&-(8\alpha+3\beta)\mathcal{Q}_{\mu\nu,\alpha\beta}\,,
\eea
where
\bea
\label{2.16}
\mathcal{Q}_{\mu\nu,\alpha\beta}\Longrightarrow\begin{cases}\mathcal{Q}_{ij,kl}=\bar{\mathcal{P}}^0_{ij,kl}\,,\\
\mathcal{Q}_{00,00}=\frac{\vk^4}{k^4}\,,\\
\mathcal{Q}_{00,mn}=-\frac{\vk^2 k_mk_n}{k^4}=\mathcal{Q}_{mn,00}\,,\\
0,\quad\mathrm{otherwise}.\end{cases}
\eea

The  propagator in (\ref{2.15}) has two poles besides the term $\mathcal{Q}$.  The first line of that equation is just what one gets from pure GR, and corresponds to the massless pole $-k^2=\omega^2-\vk^2=0$. By looking at the second line in (\ref{2.15}), we notice that this pole,  for nonvanishing $\beta$, gains a correction that is proportional to $P^2$ in its spatial indices sector. In the  next section we will analyze the contribution of this correction to the dynamic of the theory, at the tree-level. Besides this just mentioned pole, there is a new pole corresponding to a massless spin 2 particle which obeys a nonrelativistic dispersion relation
\be
\label{2.17}
\omega^2=\vk^2(1-\beta\vk^2)\,.
\ee
For this pole to have physical meaning either  $\beta<0$, or, otherwise, there will be a limit in the particle momentum ($\beta>0\Rightarrow\vk^2<1/\beta$). This is the expected pole in the propagator we wrote in (\ref{1}) that may improve renormalizability. Nevertheless, the last term in (\ref{2.15}), proportional to  $\mathcal{Q}$,  clearly spoils renormalizability, unless we set $8\alpha+3\beta=0$.
Such choice of the parameters $\alpha$ and $\beta$ is in accordance with the detailed balance condition introduced by Ho\v rava in \cite{Horava2}. This can be seen by taking $\alpha=-\frac38\beta$ so that the extra spacial derivative terms in the action (\ref{2}) furnishes
\bea
\label{2.18}
\int d^4x \sqrt{-g}\left(\alpha R^{(3)2}+\beta R^{(3)ij}R^{(3)}_{ij}\right)
&=&\beta\int d^4x \sqrt{-g}\left(-\frac{3}{8} R^{(3)2}+ R^{(3)ij}R^{(3)}_{ij}\right)\nonumber\\
&=&\beta\int d^4x\sqrt{-g}\frac{\delta W[q]}{\delta q_{ij}}\mathcal{G}_{ij,kl}\frac{\delta W[q]}{\delta q_{kl}}\,,
\eea
where
\be
\label{2.19}
\mathcal{G}_{ij,kl}=q_{i(k}q_{l)j}-\frac{1}{2}q_{ij}q_{kl}
\ee
is the inverse of the Weyl metric $\mathcal{G}^{ij,kl}=q^{i(k}q^{l)j}-q^{ij}q^{kl}$, $q_{ij}=g_{ij}$ while, $q^{ij}$ is the inverse of $q_{ij}$ and the 3-dimensional action $W$ is given by
\be
\label{2.20}
W[q]=\int_\sigma d^3x \sqrt{q}R^{(3)}\,.
\ee
It is worth mentioning that the detailed balance condition plays an essencial role regarding renormalizability in the full Ho\v rava theory, as shown in \cite{Orlando}. In the present case it justifies the removal of a bad behaved term which would spoil the renormalization of the model. 

\section{Tree-level unitarity}
\label{sec:3}

We can readily check from (\ref{2.15}) that the linearized theory described by (\ref{2}) does not contain tachyons. This is already an improvement compared to the modified theories of gravity with higher time derivatives~\cite{Stelle,Accioly} that, beyond having tachyons, are not unitary even at the tree-level.

In this section we are going to examine unitarity for the model (\ref{2}) at the tree-level. To this end, it is enough to study the residue of the poles when we saturate the propagator (\ref{2.15}) with an arbitrary conserved current~\cite{Nieuwenhuizen,Accioly}
\be
\label{3.1}
T^{\mu\nu}=ak^\mu k^\nu + b\tilde{k}^\mu\tilde{k}^\nu+c_{xy}\ep^\mu_x\ep^\nu_y+2d k^{(\mu}\tilde{k}^{\nu)}+2e_x\ep_x^{(\mu}k^{\nu)}+2f_x\ep^{(\mu}_x\tilde{k}^{\nu)}\,.
\ee
Here, $T^{\mu\nu}$ has been  arbitrarily expanded in terms of the linearly independent four-vectors $k^{\mu}=(\omega,\vk)$, $\tilde{k}^\mu=(-\omega,\vk)$ and the orthonormal graviton polarization vectors $\ep^\mu_x=(0,\vec{\ep}_x)$, $x,y=1,2$ ($\vec{\ep}_x\cdot\vk=0$) with the coefficients $a$, $b$, $c_{xy}=c_{yx}$, $d$, $e_x$ and $f_x$. Conservation implies $T^{\mu\nu}k_\nu=0$, so that
\bml
\label{3.2}
\bea
&&ak^2+d(\omega^2+\vk^2)=0\,,\label{3.2a}\\
&&dk^2+b(\omega^2+\vk^2)=0\,,\label{3.2b}\\
&&e_xk^2+f_x(\omega^2+\vk^2)=0\,,\label{3.2c}\\
&&ak^4+b(\omega^2+\vk^2)^2+2dk^2(\omega^2+\vk^2)=0\,.\label{3.2d}
\eea
\eml
Equations (\ref{3.2}) and (\ref{3.1}) enable us to write the amplitude
\be
\label{3.3}
T^{\mu\nu}\opt^{-1}_{\mu\nu,\alpha\beta}T^{\alpha\beta}=\frac{\left(c^{11}-c^{22}\right)^2+4(c^{12})^2}{-k^2+\beta\vk^4}-16(8\alpha+3\beta)k^4 a^2\frac{\left(k^4-3k^2\vk^2+3\vk^4\right)^2}{\left(\omega^2+\vk^2\right)^2}\,,
\ee
where we used the identities $T\mathcal{P}T=0$ for $\mathcal{P}=\mathcal{P}^1,\bar{\mathcal{P}}^0, \bar{\bar{\mathcal{P}}}^0$ and
\bea
\label{3.4}
T^{ij}2P_{ij,kl}T^{kl}&=&T^{\mu\nu}\left(2\mathcal{P}^2-\mathcal{P}^0\right)_{\mu\nu,\alpha\beta}T^{\alpha\beta}\nonumber\\
&=&\left(c^{11}-c^{22}\right)^2+4(c^{12})^2\ge0\,.
\eea

The last term in (\ref{3.3}) does not correspond to a pole, so, there is no particle associated to it. But, it clearly prejudices the divergent behavior in perturbation theory unless we fix accordingly the constants $\alpha$ and $\beta$, i.e., $8\alpha+3\beta=0$.
In this situation, we conclude that, at least at the tree-level, the theory is unitary. The cancellation of the term corresponding to  the pole at $-k^2=0$ in (\ref{3.3}) implies that this particle has no dynamics. Yet, the pole at $-k^2+\beta\vk^2=0$ with positive residue is a physical particle with two degrees of freedom ($P^2_{ij,kl}$ has only two independent indices). In other words, apparently, the disappearance of the pole $-k^2=0$ in favor of the pole $-k^2+\beta\vk^2=0$ means that the graviton has become a nonrelativistic particle represented by this pole. Notice that the residue of this modified pole is the same as the graviton pole when $\alpha=\beta=0$.

\section{Semiclassical potential in the nonrelativistic limit}
\label{sec:4}

\begin{figure}
\includegraphics[scale=.5,angle=-90]{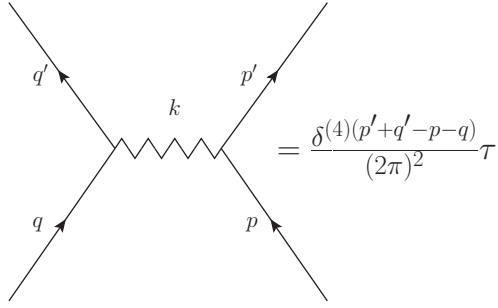}
\caption{\label{Fig1}Second order Feynman diagram of two bosons interacting via gravitational exchange.}
\end{figure}

\begin{figure}
\includegraphics[scale=.5,angle=-90]{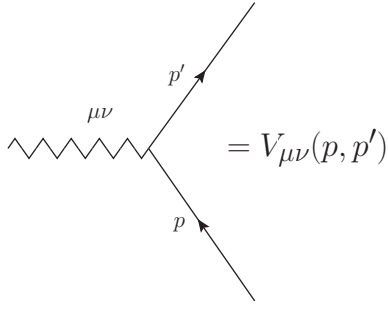}
\caption{\label{Fig2}Boson-boson interaction vertex.}
\end{figure}

Once the propagator of the theory is obtained, we can analyze the scattering process of two particles interacting gravitationally within this modified theory. This enable us to evaluate the effective low energy potential due to the gravitational interaction of two identical massive bosons  particles of zero spin  described by the Lagrangian density
\bea
\label{4.1}
\mathcal{L}&=&\sqrt{-g}\left(-\partial_\mu\varphi\partial^\mu\varphi^*-m^2\varphi\varphi^*\right)\nonumber\\
&\approx&\mathcal{L}_{\mathrm matter}+\mathcal{L}_I\,,
\eea
up to first order in $h$, where
\bml
\label{4.2}
\bea
\mathcal{L}_{\mathrm matter}&=&-\partial_\mu\varphi\partial^\mu\varphi^*-m^2\varphi\varphi^*\,,\label{4.2a}\\
\mathcal{L}_I&=&-\frac{\kappa}{2} h^{\mu\nu}\underbrace{\left[2\partial_{(\mu}\varphi\partial_{\nu)}\varphi^*-\eta_{\mu\nu}\left(\partial_\alpha\varphi\partial^\alpha\varphi^*
+m^2\varphi\varphi^*\right)\right]}_{T_{\mu\nu}}\,.\label{4.2b}
\eea
\eml
The amplitude for  this scattering process, as  illustrated in Fig.~\ref{Fig1}, is given by

\be
\label{4.3}
\tau\equiv V^{\mu\nu}(q,q^\prime)\opt^{-1}_{\mu\nu,\alpha\beta}(k)V^{\alpha\beta}(p,p^\prime)\,,
\ee
where $k=p^\prime-p=q-q^\prime$ is the momentum transfer and the vertex $V_{\mu\nu}(p,p^\prime)$, drawn in Fig.~\ref{Fig2}, is
\be
\label{4.4}
V_{\mu\nu}(p,p^\prime)\equiv-\frac{\kappa}{2}\left[2p_{(\mu}p^\prime_{\nu)}-\eta_{\mu\nu}\left(p\cdot p^\prime-m^2\right)\right]\,,
\ee
with $p\cdot p^\prime\equiv p^\mu p_\mu^\prime$.
For simplicity, we will restrict our calculation to  the center of momentum reference frame ($p^\mu=(E,\vp)$, $p^{\prime\,\mu}=(E,\vp^{\,\prime})$, $q^\mu=(E,-\vp)$, and $q^{\prime \,\mu}=(E,-\vp^{\,\prime})$), such that the amplitude (\ref{4.3}) can be cast as
\bea
\label{4.5}
\tau&=&-\frac{\kappa^2}{2\vk^2}\left[\left(q\cdot p+q\cdot p^\prime\right)\left(q^\prime\cdot p+q^\prime\cdot p^\prime\right)
-\left(q\cdot q^\prime+m^2\right)\left(p\cdot p^\prime+m^2\right)-2m^4\right]\nonumber\\
&&+\frac{\kappa^2}{2}\left[\left(\vq\cdot \vp+\vq\cdot \vp^{\phantom a\prime}\right)\left(\vq^{\phantom a\prime}\cdot \vp+\vq^{\phantom a\prime}\cdot \vp^{\phantom a\prime}\right)-\frac{1}{2}\vq\cdot(\vq+\vq^{\phantom a \prime})\vp\cdot(\vp+\vp^{\phantom a\prime})\right]\left(\frac{1}{\vk^2}-\frac{1}{\vk^2-\beta\vk^4}\right)\nonumber\\
&&-(8\alpha+3\beta)\frac{\kappa^2}{4}\left(2p^0p^{0\prime}+p\cdot p^\prime+m^2\right)^2\nonumber\\
&&=-\frac{\kappa^2}{2\vk^2}\left[\left(q\cdot p+q\cdot p^\prime\right)\left(q^\prime\cdot p+q^\prime\cdot p^\prime\right)
-\left(q\cdot q^\prime+m^2\right)\left(p\cdot p^\prime+m^2\right)-2m^4\right]\nonumber\\
&&+\frac{\kappa^2}{4}\left(p\cdot q^\prime+m^2\right)^2\left(\frac{1}{\vk^2}-\frac{1}{\vk^2-\beta\vk^4}\right)-(8\alpha+3\beta)\frac{\kappa^2}{4}\left(2E^2+p\cdot p^\prime+m^2\right)^2\,.
\eea

The effective potencial for the just calculated scattering amplitude is obtained from the Fourier transform
\be
\label{4.6}
V(\vx)=\frac{1}{8m^2}\int \frac{d^3k}{(2\pi)^3}\tau_{NR}\exp(-i\vk\cdot\vx)\,,
\ee
by inserting  the low energy limit of (\ref{4.5}), that is to say,
\be
\label{4.7}
\tau_{NR}=-\frac{\kappa^2m^4}{\vk^2}-(8\alpha+3\beta)\kappa^2m^4\,.
\ee
Collecting all the above results one gets
\be
\label{4.8}
V(\vx)=-\frac{Gm^2}{|\vx|}-2(8\alpha+3\beta)Gm^2\delta^{(3)}(\vx)\,.
\ee
As one can observe, the nonrelativistic limit (\ref{4.7}) eliminates the pole $-k^2+\beta\vk^4=0$. The pure GR sector obviously reproduces Newton's potential. But this is not all. The extra non-pole term in (\ref{2.15}) is the responsible for the appearance of a Dirac's delta in the potential. This is similar to what happens in QED in the calculation of the nonrelativistic effective potential of two electrons exchanging a photon. In that case, this extra delta term, also known as Darwin's term, is interpreted as quantum fluctuations in the electron's position due to its position indeterminancy.

\section{Summary and conclusions}
\label{Conclusion}

In this paper we have studied the propagator of a Hora\v va-Lifshitz like theory with quartic spatial derivative terms. This  propagator has two poles, one corresponding to the GR graviton pole and other  also corresponding  to a spin 2 massless but of nonrelativistic character.
Besides the improvement brought by the spatial quartic terms there are some points which deserve mentioning.
Firstly, the presence of the unmodified GR sector is potentially dangerous. Nevertheless,  by coupling the model to a scalar field through a conserved current, we verified that at tree-level the residue of the GR pole is zero
and that  the only excitation corresponds to a spin 2 nonrelativistic particle.
In that situation,  we also showed that  the theory is free of ghosts and tachyons.
 Even though at the tree-level the unmodified GR pole has no dynamics, we cannot assert that it will not contribute to higher order virtual processes. Moreover, the propagator also possess a non-pole term that by power counting clearly prejudices renormalizability . In fact, such term increases the  ultraviolet divergence and may spoil the theory as a whole. The solution for this problem passes by  a choice of the arbitrarily inserted constants $\alpha$ and $\beta$ and, as showed, is in the class of extensions which satisfies the detailed balance condition.

The effective low energy potential  for scattering process involving two massive bosons that interact via this higher spatial derivative theory was also computed  and, as a result, we obtained Newton's potential plus a Dirac delta in position, i.e., a Darwin-like term. This term is well known in QED from the evaluation of the nonrelativistic potential obtained from electron-electron scattering and is related to the indeterminancy in electrons' position. In the quantum gravity case, this imprecision in the boson position is not present in pure linearized GR and have appeared in the present case labeled by the constants $\alpha$ and $\beta$.

The relevant information we have obtained with this model is that higher order spatial derivatives do not ensure that the GR propagator will be modified as expected, with higher order in momentum $\vk^2$. In fact, within the modifications worked in this paper, we have reobtained the unwished GR graviton pole plus the desired term like (\ref{1}) which may improve the ultraviolet behavior of the quantum theory. We also showed that a bad ultraviolet behaving term has appeared and that only within a specific combination of the constants $\alpha$ and $\beta$ it can be eliminated. This, in fact, shows that the addition of higher order spatial derivatives is not enough to warrant renormalizability and that the way one introduces such extra terms is crucial  when renormalizability is at stake.

\begin{acknowledgments}
The authors thank M. Dias and Pedro R. S. Gomes for useful discussions. This work was partially supported by Fundação de Amparo à Pesquisa do Estado de São Paulo (FAPESP) and Conselho Nacional de Pesquisas (CNPq).
\end{acknowledgments}

\appendix

\section{ Barnes-Rivers operators}
\label{ApendiceA}

The $3$-dimensional symmetric Barnes-Rivers operators~\cite{Rivers,Nieuwenhuizen,Accioly} are given by
\bml
\label{23}
\bea
P^1_{ij,kl}&=&2\theta_{((i(k}\omega_{l)j))}\,,\label{23a}\\
P^2_{ij,kl}&=&\theta_{i(k}\theta_{l)j}-\frac{1}{2}\theta_{ij}\theta_{kl}\,,\label{23b}\\
P^0_{ij,kl}&=&\frac{1}{2}\theta_{ij}\theta_{kl}\,,\label{23c}\\
\bar{P}^0_{ij,kl}&=&\omega_{ij}\omega_{kl}\,,\label{23d}\\
\bar{\bar{P}}^0_{ij,kl}&=&\theta_{ij}\omega_{kl}+\omega_{ij}\theta_{kl}\,,\label{23e}
\eea
\eml
where the projection tensors
\be
\label{24}
\theta_{ij}=\delta_{ij}-\frac{k_i k_j}{\vk^2}\,,\qquad \omega_{ij}=\frac{k_i k_j}{\vk^2}
\ee
have been defined. Such operators obey (using $AB$ in the place of $A^{ij,kl}B_{kl,mn}$ to the contractions) $P^1P^1=P^1$, $P^2P^2=P^2$, $P^0P^0=P^0$, $\bar{P}^0\bar{P}^0=\bar{P}^0$, $\bar{\bar{P}}^0\bar{\bar{P}}^0=(D-1)(P^0+\bar{P}^0)$, $P^0\bar{\bar{P}}^0=\bar{\bar{P}}^0\bar{P}^0=P^{\theta\omega}$, $\bar{P}^0\bar{\bar{P}}^0=\bar{\bar{P}}^0P^0=P^{\omega\theta}$, together with $P^{\theta\omega}_{ij,kl}=\theta_{ij}\omega_{kl}$ and $P^{\omega\theta}_{ij,kl}=\omega_{ij}\theta_{kl}$. Any other contraction is found to be zero. Those operators also obey the identities
\bml

\label{25}
\bea
&&\delta_{ij,kl}=(P^1+P^2+P^0+\bar{P}^0)_{ij,kl}\,\label{25a}\\
&&\delta_{ij}\delta_{kl}=(2P^0+\bar{P}^0+\bar{\bar{P}}^0)_{ij,kl}\,,\label{25b}\\
&&\frac{4}{\vk^2}\delta_{((i(k}k_{l)}k_{j))}=(2P^1+4\bar{P}^0)_{ij,kl}\,,\label{25c}\\
&&\frac{1}{\vk^2}(\delta_{ij}k_k k_l+\delta_{kl}k_i k_l)=(\bar{\bar{P}}^0+2\bar{P}^0)_{ij,kl}\,,\label{25d}\\
&&\frac{1}{\vk^4}(k_i k_j k_k k_l)=\bar{P}^0_{ij,kl}\,.\label{25e}
\eea
\eml
The above tools enable one to write an arbitrary symmetric operator $\opt$ as
\be
\label{25-1}
\opt=x_1P^1+x_2P^2+x_0P^0+\bar{x}_0\bar{P}^0+\bar{\bar{x}}_0\bar{\bar{P}}^0\,,
\ee
whose inverse, if exists, will be
\be
\label{25-2}
\opt^{-1}=\frac{P^1}{x_1}+\frac{P^2}{x_2}+\frac{1}{x_0\bar{x}_0-2\bar{\bar{x}}_0^2}\left(\bar{x}_0P^0+x_0\bar{P}^0-\bar{\bar{x}}_0\bar{\bar{P}}^0\right)\,.
\ee

In $4$-dimensions, the Barnes-Rivers operators may be written as
\bml
\label{26}
\bea
\mathcal{P}^1_{\mu\nu,\alpha\beta}&=&2\Theta_{((\mu(\alpha}\Omega_{\beta)\nu))}\,,\label{26a}\\
\mathcal{P}^2_{\mu\nu,\alpha\beta}&=&\Theta_{\mu(\alpha}\Theta_{\beta)\nu}-\frac{1}{3}\Theta_{\mu\nu}\Theta_{\alpha\beta}\,,\label{26b}\\
\mathcal{P}^0_{\mu\nu,\alpha\beta}&=&\frac{1}{3}\Theta_{\mu\nu}\Theta_{\alpha\beta}\,,\label{26c}\\
\bar{\mathcal{P}}^0_{\mu\nu,\alpha\beta}&=&\Omega_{\mu\nu}\Omega_{\alpha\beta}\,,\label{26d}\\
\bar{\bar{\mathcal{P}}}^0_{\mu\nu,\alpha\beta}&=&\Theta_{\mu\nu}\Omega_{\alpha\beta}+\Omega_{\mu\nu}\Theta_{\alpha\beta}\,.\label{26e}
\eea
\eml
Now, the projection operators are defined by
\be
\label{27}
\Theta_{\mu\nu}=\delta_{\mu\nu}-\frac{k_\mu k_\nu}{k^2}\,,\qquad \Omega_{\mu\nu}=\frac{k_\mu k_\nu}{k^2}\,.
\ee

Eventually, it will be convenient to relate the Barnes-Rivers operators in three and four dimensions. When only the spatial indices are being treated, it is possible to obtain the identities between the $P$'s and $\mathcal{P}$'s as
\bml
\label{28}
\bea
P^2_{ij,kl}&=&\left[\mathcal{P}^2+\frac{\omega^2}{\vk^2}\mathcal{P}^1-\frac{1}{2}\mathcal{P}^0+\frac{\omega^4}{2\vk^4}\bar{\mathcal{P}}^0
-\frac{\omega^2}{2\vk^2}\bar{\bar{\mathcal{P}}}^0\right]_{ij,kl}\,,\label{28a}\\
P^1_{ij,kl}&=&\left(\frac{k^2}{\vk^2}\mathcal{P}^1+\frac{2\omega^2k^2}{\vk^4}\bar{\mathcal{P}}^0\right)_{ij,kl}\,,\label{28b}\\
P^0_{ij,kl}&=&\left(\frac{3}{2}\mathcal{P}^0+\frac{\omega^4}{2\vk^4}\bar{\mathcal{P}}^0+\frac{\omega^2}{2\vk^2}\bar{\bar{\mathcal{P}}}^0\right)_{ij,kl}\label{28c}\\
\bar{P}^0_{ij,kl}&=&\frac{k^4}{\vk^4}\bar{\mathcal{P}}^0_{ij,kl}\,,\label{28d}\\
\bar{\bar{P}}^0_{ij,kl}&=&\left(\frac{k^2}{\vk^2}\bar{\bar{\mathcal{P}}}^0+\frac{2\omega^2k^2}{\vk^4}\bar{\mathcal{P}}^0\right)_{ij,kl}\,.\label{28e}
\eea
\eml
Such equations enable one to get
\be
\label{29}
\left(2\mathcal{P}^2-\mathcal{P}^0-3\bar{\mathcal{P}}^0-\bar{\bar{\mathcal{P}}}^0\right)_{ij,kl}=\left(2P^2
-\frac{2\omega^2}{k^2}P^1+\frac{k^2-4\vk^2}{k^2}\bar{P}^0-\bar{\bar{P}}^0\right)_{ij,kl}\,.
\ee


\end{document}